\documentclass[preprint,preprintnumbers,amsmath,amssymb]{revtex4}
\usepackage{epsfig}
\usepackage{amssymb}
\usepackage{graphicx}
\usepackage{bm}

\usepackage{latexsym}
\usepackage{amsmath}
\usepackage{slashed}
\usepackage{graphics}
\usepackage{epsfig}
\def\br{\begin{eqnarray}}
\def\er{\end{eqnarray}}
\def\be{\begin{equation}}
\def\ee{\end{equation}}

\def\({\left(}
\def\){\right)}
\def\<{\left\langle}
\def\>{\right\rangle}

\begin{document}
\title{Extreme walking behavior and SDE boundary conditions in a 331-TC model}
\author{A. Doff}
\email{agomes@utfpr.edu.br}
\affiliation{Universidade Tecnol\'ogica Federal do Paran\'a - UTFPR - DAFIS
Av Monteiro Lobato Km 04, 84016-210, Ponta Grossa, PR, Brazil }

\date{\today}

\begin{abstract}
The solution of the phenomenological problems of technicolor (TC) models may reside in the different dynamical behavior of the technifermions self-energy appearing in walking (or quasi-conformal) theories. Motivated by recent results where it is shown how the boundary conditions (bc) of the  anharmonic oscillator representation of the Schwinger-Dyson gap equation (SDE) to $SU(N)$ are directly related  with  the mass anomalous dimensions, and different bc cause a change in the ultraviolet asymptotic behavior of the self-energies, in this work we verify that it is possible to have a hard technifermion self-energy in TC models originated through radiative corrections coming from the interactions mediated by the new massive neutral and charged gauge bosons, $Z'$ and  $U^{\pm\pm}$  in the context of a 331-TC model. 
\end{abstract}


\maketitle

\section{Introduction}

\par The origin of fermion and gauge boson masses in the standard model (SM) of elementary particles is explained by their interaction with the Higgs boson. 
The $125$ GeV new resonance discovered at the LHC \cite{LHC1,LHC2} has many of the characteristics expected for the Standard Model (SM) Higgs boson, however
the data still cannot discard the possibility of this boson to be a composite one. If this particle is a composite or an elementary scalar boson is still
an open question that probably may be answered by the future LHC data.

\par The case of a composite state generating dynamical symmetry breaking(DSB) instead of an elementary one is more akin to the phenomenon of spontaneous 
symmetry breaking that originated from the Ginzburg-Landau Lagrangian, which can be derived from the microscopic BCS theory of superconductivity describing the electron-hole interaction (or the composite state in our case). A similar mechanism happens in QCD where the chiral symmetry breaking is promoted by a non-trivial vacuum expectation value of a fermion bilinear operator and the Higgs role is played by the composite $\sigma$ meson. In particular, the TC idea was the earliest attempt of building models in this direction \cite{tca,tcc,tcd}.

\par  In some  extensions of the standard model (SM), as in the so called 3-3-1 models\cite{331a,331b,331c,331d,331e} $SU(3)_{{}_{L}}\otimes SU(3)_{{}_{c}} \otimes U(1)_{{}_{X}}$,  new massive neutral and charged gauge bosons, $Z'$  and  $V^{\pm}, U^{\pm\pm} $, are predicted.   The 3-3-1 model is the minimal gauge group that at the leptonic level admits charged fermions and their antiparticles as members of the same multiplet,  the extra predictions of these $GW = SU(2)_{{}_{L}}\otimes U(1)_{{}_{Y}}$ alternative models are leptoquark fermions with electric charges $-5/3$ and $4/3$ and bilepton gauge bosons with lepton number $L = \pm 2$. The quantization of electric charge is inevitable in the $G=3-m-1$(to m=3,4) models\cite{QQ1, QQ2, QQ3, QQ4, QQ5, QQ6, QQ7,QQ8} with three non-repetitive fermion generations, breaking universality independently on the character of the neutral fermions.

\par  In the Refs.\cite{Das,331x} it was suggested that the gauge symmetry breaking of a specific version of a 3-3-1 model\cite{331x} could be implemented dynamically, because at the scale of a few TeVs the $U(1)_X$ coupling constant becomes strong and the  exotic quark $T$(charge $5/3$)  will form a  $U(1)_X$ condensate breaking  $SU(3)_{{}_{L}}\otimes U(1)_X$ to the electroweak symmetry.  This possibility was explored in the Refs.\cite{331x,331y,331z,331w} assuming a model based on the gauge symmetry  $SU(2)_{TC}\otimes SU(3)_{{}_{L}}\otimes SU(3)_{{}_{c}} \otimes U(1)_{{}_{X}}$(331-TC model), where the electroweak symmetry is broken dynamically by a technifermion condensate, that is characterized by the $SU(2)_{TC}$ Technicolor (TC) gauge group.  

\par The early technicolor models \cite{tca,tcb,tcc,tcd,tce}  suffered from problems like flavor changing neutral currents (FCNC) and contributions to the electroweak corrections not compatible with the experimental data, as can be seen in the reviews of Ref.\cite{tc1,tc2,tc3,tc4}. However, the TC dynamics may be quite different from the known strong interaction theory, i.e. QCD, this fact has led to the walking TC proposal \cite{walk1,walk2,walk3,walk4,walk5,walk6,walk7,walk8,walk9}, which are theories where the incompatibility with the experimental data has been solved, making the new strong interaction almost conformal and changing appreciably its dynamical behavior. 

It is  possible to obtain an almost conformal TC theory when the fermions are in the fundamental representation, with the introduction of a large number of TC fermions ($n_{TF}$), leading to an almost zero $\beta$ function and flat asymptotic coupling constant. The cost of such procedure may be a large S parameter\cite{peskin1,peskin2}  incompatible with the high precision electroweak measurement when it is assumed that technicolor is just QCD scaled up to a higher energy scale \cite{lane2}. However, this problem can be solved by assuming that TC fermions are in other representations than the fundamental one \cite{sannino1,sannino2,sannino3}, and an effective Lagrangian analysis indicates that such models also imply in a light scalar Higgs boson \cite{sannino3}. This possibility was also investigated and confirmed  through the use of an effective potential for composite operators \cite{us1} and through a calculation involving the Bethe-Salpeter equation (BSE) for the scalar state \cite{us2}.

\par In Refs.\cite{us1,us2,us3,twoscale} we discussed the possibility of  obtaining a  light composite TC scalar boson,  this result is a direct consequence of an extreme walking (or quasi-conformal) technicolor theories, where the  asymptotic TC self-energy  behavior  is described by the so called irregular form. There are  different ways to obtain  of extreme walking (or quasi-conformal) behavior in technicolor theories,  in the Ref.\cite{331-4f}  this behavior was obtained based on a 331-TC model. The present work was motivated by the  results obtained in Ref.\cite{us4}, where it is discussed how the boundary conditions of the  anharmonic oscillator representation of the Schwinger-Dyson gap equation (SDE) for $SU(N)$ gauge theories are directly related  with  the mass anomalous dimensions,  and the result of Ref.\cite{ardn} where it was verified that when TC is embedded into a larger theory including also QCD, the radiative corrections, coupling the different strongly interacting Dyson equations, change the ultraviolet behavior of the gap equation solution. Here we show  that it is possible to have a hard technifermion self-energy in TC models originated through radiative corrections coming from the interactions mediated by the new massive neutral and charged gauge bosons, $Z'$ and  $U^{\pm\pm}$,  in the context of the 331-TC model. 

\par This  article is organized as follows: In section II we present the  TC Schwinger-Dyson equations (SDE) for exotic techniquarks, including  $U(1)_{X}$  corrections such  that  exotic techniquarks are coupled to  themselves via exchange of a $Z'$ boson. The inclusion of these corrections change the boundary conditions of the $Q'=(U',D')$ TC gap equations and the asymptotic behavior of the self-energy. In section III we extend this result to the usual techniquarks sector, $Q=(U,D)$, and  we show that the inclusion of $U^{\pm\pm}$  corrections will change the boundary conditions of the $Q=(U,D)$ TC gap equations leading to a quasi-conformal behavior for  all technicolor self-energies ($\Sigma_{Q'}(p)$ and $\Sigma_{Q}(p)$) resulting in a totally different model of dynamical gauge symmetry breaking. In Section V we draw our conclusions.  

\section{The $U(1)_{X}$ corrections to the $(U',D')$ TC  gap equation}

\par In the Ref.\cite{331-4f}  we computed the dynamically generated masses of heavy exotic quarks $(T,S,D)$ and  exotic techniquarks  $Q'=(U', D')$, and reproduced the results obtained in Ref.\cite{Das} for heavy  exotic quarks. The TC SDE including  $U(1)_{X}$ corrections is depicted in Fig.(1), where the curly line correspond to technigluons and the wavy line to $Z'$ bosons 
\begin{figure}[t]
\centering
\includegraphics[width=0.8\columnwidth]{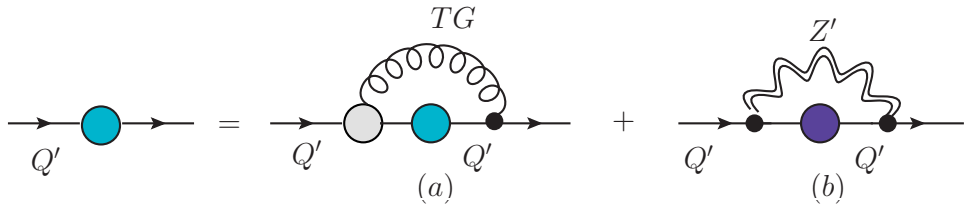}
\caption[dummy0]{The Schwinger-Dyson equation for  TC including  $U(1)_{X}$  corrections }
\label{fig1}
\end{figure}

In the sequence we will consider the TC SDE for exotic techniquarks (see Ref.\cite{331-4f}) including  the $U(1)_{X}$  correction such  that  exotic techniquarks are coupled to themselves via exchange of a $Z'$ boson (Fig.(1b)). Our main point is to show that the inclusion of this
type of correction change the boundary conditions of the  TC gap equations as emphasized recently in Ref.\cite{us4}, where it was shown how the asymptotic solution of the gap equation is modified by this new condition. Including the $U(1)_{X}$ correction depicted in Fig.(1) into the
SDE, where the self-energy, coupling constant and respective Casimir operator are indicated by the index $(TC)$, and the $Z'$ boson exchange is indicated by the index $(Z')$ we obtain following gap equation  
\br 
&&\Sigma_{Q'}(p^2)=\frac{3C_{{}_{TC}}\alpha_{{}_{TC}}}{4\pi}\left(\frac{1}{p^2 + m^2_{G}}\int^{p^2}\hspace*{-0.3cm} dk^2\frac{k^2\Sigma_{Q'}(k^2)}{k^2+\Sigma^2_{Q'}(k^2)} + \int^{\Lambda^2}_{p^2}\hspace*{-0.3cm} dk^2\frac{\Sigma_{Q'}(k^2)}{k^2 + m^2_{G}}\frac{k^2}{k^2+\Sigma^2_{Q'}(k^2)}\right)  \nonumber \\
&& + \frac{3\alpha_{{}_{X}}(Y_{{}_{L}}Y_{{}_{R}})}{4\pi}\left(\frac{1}{p^2 + M^2_{Z'}} \int_{0}^{p^2}\hspace*{-0.3cm}d^2k \frac{k^2 \Sigma_{Q'}(k^2)}{k^2 + \Sigma^2_{Q'}(k^2)} +\int_{p^2}^{\Lambda^2}\hspace*{-0.3cm} d^2k \frac{k^2\Sigma_{Q'}(k^2)}{k^2 + \Sigma^2_{Q'}(k^2)}\frac{1}{k^2 + M^2_{Z'}}\right),
\label{e1}
\er  
\noindent where $M_{Z'}$ is the $Z'$ boson mass, 
 $Y_{L,R}$ are  $U(1)_{{}_{X}}$ hypercharges attributed to the chiral components of the exotic techniquarks (Q')\cite{331-4f}. To obtain the last equation  we  also assumed the angle approximation\cite{Craig} to transform the terms  $\frac{1}{(p - k)^2 + M^2_{Z'}}$ as
\br 
\frac{1}{(p - k)^2 + M^2_{Z'}} = \frac{\theta(p - k)}{p^2 + M^2_{Z'}} + \frac{\theta(k - p)}{k^2 +M^2_{Z'}}.
\label{e2} 
\er 

In order to illustrate how the boundary conditions of the  exotic techniquarks(Q') TC gap equation are modified  by the $U(1)_{X}$ correction,
we can assume a compact notation where $(1=TC)$ and $ (2=U(1)_X)$ such that 
\br  
&&\hspace*{-0.4cm} f_1(x) = \sigma_1(x)I^a_1(x) + \theta_1I^b_1(x) + \sigma_2(x)I^a_2(x) + \,\theta_2I^b_2(x)  
\label{e3}
\er
\noindent where we introduced the following set of new variables and auxiliary functions
\br 
&& p^2 = M^2_{Q'}x \,\,\,,\,\,\, \Sigma_i(p^2) = M^2_{Q'} f_i(x)\,\,\,,\,\,\, \omega_{1} =\frac{m^2_{{}_{G}}}{M^2_{Q'}}\nonumber \\
&& k^2 = M^2_{Q'}y \,\,\,,\,\,\,\Sigma_i(k^2) = M^2_{Q'} f_i(y)\,\,\,,\,\,\,  \omega_{2}=\frac{M^2_{{}_{Z'}}}{M^2_{Q'}}\nonumber \\
&&\sigma_i(x) = \frac{\theta_i}{x + \omega_i}\,\,\,,\,\,\,\theta_1 =\frac{3C_{{}_{TC}}\alpha_{{}_{TC}}}{4\pi}\,\,\,,\,\,\,\theta_2 =\frac{3\alpha_{{}_{X}}(Y_{{}_{L}}Y_{{}_{R}})}{4\pi}\nonumber \\
&& I^a_i(x) = \int^{x}_{0}dy\frac{yf_i(y)}{y+f^2_i(y)}\,\,\,,\,\,\,g_i(x) = \frac{xf_i(x)}{x+f^2_i(x)}  \nonumber \\ 
&& I^b_1(x) = \int^{\frac{\Lambda^2}{M^2_{Q'}}}_{x} dy\frac{yf_1(y)}{y+f^2_1(y)}\frac{1}{y +\omega_1} \nonumber \\
&& I^b_{2}(x) = \int^{\frac{\Lambda^2}{M^2_{Q'}}}_{x} dy\frac{yf_2(y)}{y+f^2_2(y)}\frac{1}{y + \omega_2}, \nonumber \\
\label{e4}
\er 
\noindent as mentioned before the index  $i =1({\rm or }\,2)$ denote the $TC({\rm or }\, U(1)_{X})$ contribution to the SDE, 
$m_{{}_{G}}$ represents the technigluons mass scale ($m_{{}_{G}}\sim O(\mu_{{}_{TC}})$) \cite{331c,331-4f} which was introduced in order to regularize the (IR) divergence in the gap equation. 
  
\par With Eq.(\ref{e3}) we can obtain the following relation between $f_1(x)$ and $f'_1(x)$ in the asymptotic region
 $( x = \frac{\Lambda^2}{M^2_{Q'}} >> 1)$
\br 
&& f_1(x)  = \frac{\sigma_1(x)}{\sigma'_1(x)}f'_1(x)  + \theta_2\frac{M^2_{Q'}}{\Lambda^2}I^a_2(x) \nonumber \\
\label{e5}
\er
\noindent Considering this relation and the definitions of variables and auxiliary functions presented in (\ref{e4}) it is possible to write 
\br 
&&\Sigma_1(p^2) + p^2\Sigma'_1(p^2) \approx \frac{\kappa_{{}_{X}}}{\Lambda^2}\int^{\Lambda^2}\frac{dk^2 k^2 \Sigma_2(k^2)}{k^2 + \Sigma^2_2(k^2)}, \nonumber \\ 
\label{e6}
\er
\noindent from where we can read the ultraviolet (UV) boundary condition satisfied by the $f_1(x)$ differential equation. Notice that the term due to the $U(1)_{X}$ interaction on the right side of Eq.(\ref{e6}) looks like as a bare mass being added to the TC $(Q')$ gap equation,
which can be  described by an effective four fermion interaction with an effective coupling  
\be
\kappa_{{}_{X}} = \frac{3\alpha_{{}_{X}}(Y_{{}_{L}}Y_{{}_{R}})}{4\pi} \approx O(1).
\label{e7}
\ee
It is opportune to remember that the effective four-fermion couplings, here indicated by the letter $\kappa$, are the
ones that once introduced into the SDE generate the same effect of the radiative corrections discussed in this work.

\par
\begin{figure}[t]
\begin{center}
\hspace*{2cm}\includegraphics[width=0.8\columnwidth]{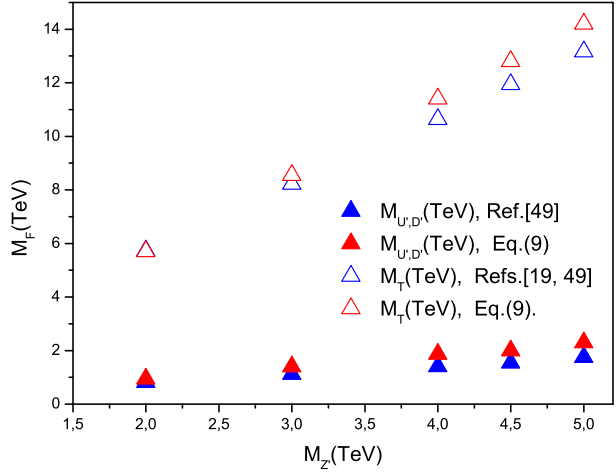}
\vspace*{-2cm}
\caption{Dynamically generated  masses $(M_F(TeV))$ of exotic quarks$(T)$ and techniquarks $Q'=(U',D')$ due to the $U(1)_{X}$ 
contribution\cite{Das, 331-4f} in comparison with the
ones obtained from the  Eq.(\ref{e9}). }
\label{fig2}
\end{center}
\end{figure}

 In Fig.(\ref{fig2})  we  show the results obtained for the dynamically generated masses due to the $U(1)_{X}$ interaction for the heavy  exotic quark $(T)$ and exotic techniquarks $Q=(U', D')$, obtained in the Ref.\cite{331-4f} and  we compare  that result with ($M_F(TeV)$) generated by the four fermion interaction described by Eq.(\ref{e8})
\br
M_F = \frac{\kappa_{{}_{X}}}{\Lambda^2}\int^{\Lambda^2}\frac{dk^2 k^2 \Sigma_2(k^2)}{k^2 + \Sigma^2_2(k^2)},
\label{e8}
\er
\noindent where $M_F = \Sigma_2(0) \approx {\rm const.}$ and we have 

\br 
M_{F}(\kappa_{{}_{X}} - 1) = \kappa_{{}_{X}}\frac{M^3_{F}}{\Lambda^2}\ln\left(\frac{\Lambda^2}{M^2_{F}}\right). 
\label{e9}
\er  
\noindent In the above expression $(\Lambda)$ is defined as a function of $M_{Z'}$ mass according to table 2 in Ref.\cite{331-4f}. 

\par The integral equation  described  by Eq.(\ref{e3}) can be transformed into a differential equation for $f_1(x)$, as reproduced below 
\be 
f^{''}_{1}(x) + \frac{2}{x + \omega_1}f^{'}_{1}(x) - \sigma'_1(x)\frac{xf_1(x)}{(x + f^2_1(x))} = 0, 
\label{e10}
\ee 
\noindent therefore,  the $U(1)_{X}$  correction  change the boundary conditions of the TC SDE  equation for the exotic 
techniquarks $Q'=(U',D')$.  These SDE boundary conditions will also cause a change in the anomalous mass dimension ($\gamma_m$) as discussed in Ref.\cite{us4}, leading to  $\gamma_m \sim O(2)$ because $\kappa_{{}_{X}} \approx O(1)$.

\section{The $U(1)_{X}$ contributions to $(U,D)$ TC  gap equation: $U^{\pm\pm}$ induced corrections}

\par
\begin{figure}[t]
\centering
\includegraphics[width=0.5\columnwidth]{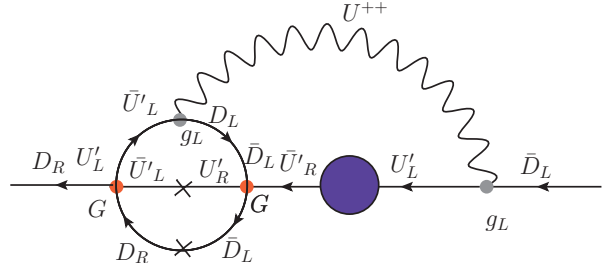}
\caption[dummy0]{$U^{++}$ boson contribution  to the $\Sigma_{D}(p^2)$  self-energy: On the left side of this figure we show  the vertex coupling  to right-handed  technifermions fields  $(D_R, U'_{R})$ induced by four fermion operators.}
\label{fig3}
\end{figure}

 In Refs.\cite{She5, She4, She3, She2, She1} the author describes in a very detailed way the possibility of $W^{\pm}$ boson coupling to right-handed fermions. These couplings are induced only at high energies, since the four fermion operators of the third fermion family will  induce a 1PI vertex function of the $W^{\pm}$ boson coupling to the right-handed fermions. In the context of the  331-TC model that we consider in this work, as shown in the previous section,  the $U(1)_{X}$ interaction  of the exotic technifermions $Q'=(U',D')$ is strong enough to generate an effective four fermion  interaction, so that at high energies the interactions mediated by bilepton bosons $U^{\pm\pm}$, containing exotic technifermions, becomes approximately vectorlike at $\mu_2 \approx \mu_{331} \approx O(TeV)$. The coupling of right-handed techniquarks $D_R$, with exotic ones $U'_{R}$,  mediated by bilepton bosons with the effective nontrivial 1PI vertex function of $U^{++}$ is given by 
\br 
{\cal L}^{eff}_{{}_{U^{++}}} & \approx &  i\frac{g^{331}_L}{\sqrt{2}}G^2N_{{}_{TC}}(Y_{{}_{L}}Y_{{}_{R}})D^{\eta}_{{}_{R}}\left\{[U'_{{}_{R}}\bar{U'}_{{}_{L}}]^{\lambda\eta}[D_{{}_{L}}\bar{D}_{{}_{R}}]^{\alpha\beta}\right.\nonumber  \\ 
&\times& \left. [U'_{{}_{L}}\bar{U'}_{{}_{L}}]^{\beta\theta} [\gamma^{\mu} P_{{}_{L}}]^{\theta\delta}[D_{{}_{L}}\bar{D}_{{}_{L}}]^{\delta\lambda}\right\}\bar{U'}^{\alpha}_{{}_{\!\!\!\!\! R}}\,U^{++}_{\mu} \nonumber \\
&  \approx & i\frac{g^{331}_L}{\sqrt{2}}D^{\eta}_{{}_{R}}[\Gamma_{++}^{\mu}(p', p)^{\eta\alpha}]\bar{U'}^{\alpha}_{{}_{\!\!\!\!\! R}}U^{++}_{\mu}(p'-p).
\er 
\noindent In the above expression the fields in brackets $[...]$ are contracted, and  $(p',p)$  are the external 
moments \cite{She1}\cite{She2}. This expression is analogous to Eqs.(28) and (29) in Ref.\cite{She1}. Considering this 
coupling in Fig.(\ref{fig3}) we show the $U^{++}$ boson  contribution to the   techniquark $D$ self-energy $\Sigma_{D}(p^2)$
(where this figure  is analogous to the Fig.4 described in Ref.\cite{She1}). 

\par
\begin{figure}[t]
\centering
\includegraphics[width=0.8\columnwidth]{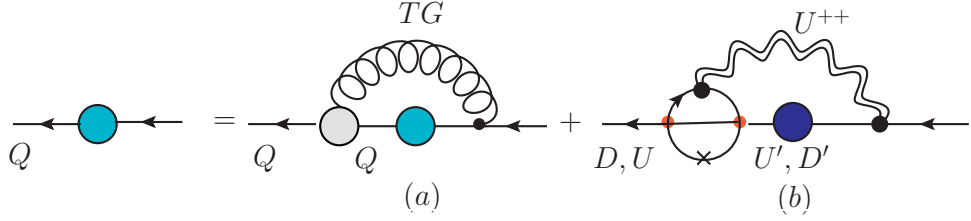}
\caption[dummy0]{The Schwinger-Dyson equation (SDE) for  techniquarks $ Q=(U,D)$ including  $U^{\pm\pm}$ bosons  corrections}
\label{fig4}
\end{figure}

 The contribution discussed above will be responsible for generating a correction to the usual techniquarks $Q=(U,D)$ SDE, as  depicted in Fig.(\ref{fig4}), where the curly lines correspond to technigluons and the wavy lines to the $U^{\pm\pm}$  correction, which can be 
identified as an bare mass being added to the TC $(Q)$ gap equation

\par Assuming the TC SDE including  the $U^{\pm\pm}$  contributions  depicted in Fig.(\ref{fig4}),  we obtain  the following SDE  satisfied by the techniquarks $ Q=(U,D)$
\br 
&&\hspace*{-0.5cm}\Sigma_{Q}(p^2)=\frac{3C_{{}_{TC}}\alpha_{{}_{TC}}}{4\pi}\left(\frac{1}{p^2 + m^2_{G}}\int^{p^2}\hspace*{-0.3cm} dk^2\frac{k^2\Sigma_{Q}(k^2)}{k^2+\Sigma^2_{Q}(k^2)}  + \int^{\Lambda^2}_{p^2}\hspace*{-0.3cm} dk^2\frac{\Sigma_{Q}(k^2)}{k^2 + m^2_{G}}\frac{k^2}{k^2+\Sigma^2_{Q}(k^2)} \hspace*{0.4cm} \right) \nonumber \\
&&\hspace*{-0.5cm} + \frac{3\alpha^{331}_{{}_{L}}}{4\pi}\frac{\gamma_{{}_{{}_{U^{\pm\pm}}}}}{\sqrt{2}} \left(\frac{1}{p^2 + M^2_{U^{\pm\pm}}} \int_{0}^{p^2}\hspace*{-0.3cm}d^2k \frac{k^2 \Sigma_{Q'}(k^2)}{k^2 + \Sigma^2_{Q'}(k^2)} + \int_{p^2}^{\Lambda^2}\hspace*{-0.3cm} d^2k \frac{k^2\Sigma_{Q'}(k^2)}{k^2 + \Sigma^2_{Q'}(k^2)}\frac{1}{k^2 + M^2_{U^{\pm\pm}}}\right)\!, 
\label{e15}
\er  
\noindent in Eq.(\ref{e15}) the factor $(\alpha^{331}_{{}_{L}}\gamma_{{}_{{}_{U^{\pm\pm}}}}/\sqrt{2})$ corresponds  to the one  obtained in Refs. \cite{She2}\cite{She1},  where $(\alpha_{{}_{W}}({\cal{E}})\gamma_{{}_{{}_{W^{\pm}}}}/\sqrt{2}\alpha_c )$, and we have assumed $\alpha_c=\frac{\pi}{3} \approx 1$. As performed in these references we can consider $\gamma_{{}_{{}_{U^{\pm\pm}}}}\!\!\! = \,(Y_{{}_{L}}Y_{{}_{R}})\Gamma_{\pm\pm}(p', p)|_{{}_{p,p' \to {\Lambda}}}  \approx  (Y_{{}_{L}}Y_{{}_{R}})$. The integral equation  described 
by Eq.(\ref{e15}) can be transformed into a differential equation  $\Sigma_{Q}(p^2)$, and  the  (UV) boundary condition that should
be satisfied  by  this  differential equation  is given by 
\br 
&&\Sigma_Q(p^2) + p^2\Sigma'_Q(p^2) \approx \frac{\kappa_{{}_{\pm\pm}}}{\Lambda^2}\int^{\Lambda^2}\frac{dk^2 k^2 \Sigma_{Q'}(k^2)}{k^2 + \Sigma^2_{Q'}(k^2)}. \nonumber \\ 
\label{e18}
\er
\noindent  The term described on the right side of Eq.(\ref{e18})  looks like as a bare mass being added to the $Q=(U,D)$  gap equation, which can be generated by an effective coupling ($\kappa_{{}_{\pm\pm}}$) 
\be
\kappa_{{}_{\pm\pm}} \approx \frac{3\alpha_{{}_{L}}}{4\pi}\frac{(Y_{{}_{L}}Y_{{}_{R}})}{\sqrt{2}} \approx  \frac{\alpha_{{}_{L}}}{\sqrt{2}\alpha_{{}_{X}}} \kappa_{{}_{X}} ,
\label{e19}
\ee 
\noindent where we defined $\alpha^{331}_{{}_{L}} = \alpha_{{}_{L}}$.  The ratio between the coupling constants 
$\frac{\alpha_{{}_{L}}}{\alpha_{{}_{X}}}$ can still be expressed in terms of the following 
relationship \cite{331a}\cite{alex1}\cite{alex2}\cite{alex3}\cite{alex4}
\begin{equation}
\frac{\alpha_{{}_{X}}}{\alpha_{{}_{L}}} = \frac{\sin^2\theta_{{}_{W}}(\mu)}{1 - 4\sin^2\theta_{{}_{W}}(\mu)}
\label{e20}
\end{equation}
\noindent where  $\theta{{}_{W}}$ is the electroweak mixing angle. In  Ref.\cite{us4}  we verified that the anomalous mass dimension can be read out directly from  the UV boundary conditions, such that, assuming the addition of an effective four fermion interaction, we have
\br 
&& \gamma_m(t_{\Lambda}) = \frac{(\gamma_m -2) + 2\kappa(1+\gamma_m)}{1+2\kappa} .
\label{e21} 
\er 
\noindent In the above equation $\gamma_m$ is the mass anomalous dimension, $\kappa$ is a four fermion effective coupling as commented 
earlier,   $t_{\Lambda} = \ln\frac{\Lambda}{\mu_{{}_{TC}}}$,  and $\mu_{{}_{TC}}$ is the characteristic TC scale. As we saw in Section II, 
$\kappa_{{}_{X}} \sim O(1)$,   and we can represent the UV  behavior of the anomalous mass dimension $\gamma_m$ attributed to
 techniquarks $Q$ assuming Eqs.(\ref{e19}), (\ref{e20}) and (\ref{e21}), which leads to $\gamma_m \approx O(0.8 - 1.3)$ assuming $\sin^2\theta_{{}_{W}}(M_Z) \approx 0.231$ as a lower limit and considering the extrapolation of the perturbative calculation presented in\cite{alex4},  as an upper limit.

\section{Conclusions}

\par    The possibility of obtaining a light composite scalar ($M_\phi \approx 125 GeV$)  along the approach discussed in Refs.\cite{us1,us2,us3,twoscale}, is a direct consequence  of extreme walking (or quasi-conformal) technicolor theories, where the  asymptotic self-energy  behavior  is described by a irregular form of the TC self-energy \cite{us1,us2}.   
Recently in Ref.\cite{331-4f} we proposed an scheme to obtain a quasi-conformal self-energy behavior based on a 331-TC model. Motivated by the result obtained  in  Ref.\cite{us4}, where we discuss how the boundary conditions of the  anharmonic oscillator representation of the 
$SU(N)$ gauge theories SDE are directly related  with  the mass anomalous dimensions. In this work we show  that it is possible to have a hard technifermion self-energy in TC models originated through radiative corrections coming from the  interactions mediated by the new massive neutral and charged gauge bosons, $Z'$ and  $U^{\pm\pm}$. 

In section II we presented the  TC SDE of exotic techniquarks, including  $U(1)_{X}$ radiative corrections, such  that  exotic techniquarks are coupled to themselves via exchange of a $Z'$ boson. We verified that the inclusion of these corrections change the boundary conditions of the $Q'=(U',D')$ TC gap equations, and we extended this result to the sector of usual techniquarks,  verifying that the inclusion of 
$U^{\pm\pm}$ radiative  corrections change the boundary conditions of the $Q=(U,D)$ TC gap equations leading to a walking 
(or quasi-conformal) behavior for $\Sigma_{Q}(p^2)$.  
 

\begin{acknowledgments}
I would like to thank A. A. Natale for reading the manuscript and for useful discussions. This research was partially supported by the Conselho Nacional de Desenvolvimento Cient\'{\i}fico e Tecnol\'ogico (CNPq)  by grant 302663/2016-9 .  
\end{acknowledgments}



\begin {thebibliography}{99}

\bibitem{LHC1} ATLAS Collaboration, Phys. Lett. B {\bf 716}, 1 (2012). 
\bibitem{LHC2} CMS Collaboration, Phys. Lett. B {\bf 716}, 30 (2012). 
\bibitem{tca} L. Susskind, Phys. Rev.  D {\bf 20} , 2619 (1979). 
\bibitem{tcc} S. Weinberg, Phys. Rev. D {\bf 13}, 974 (1976).
\bibitem{tcd} S. Weinberg, Phys. Rev. D {\bf 19} 1277 (1979). 
\bibitem{331a} M. Singer, J. W. F. Valle and J. Schechter, Phys. Rev. D {\bf  22}, 738 (1980). 
\bibitem{331b} F. Pisano and V. Pleitez, Phys. Rev. D{\bf  46}, 410 (1992).
\bibitem{331c} P. H. Frampton, Phys. Rev. Lett. {\bf 69}, 2889 (1992). 
\bibitem{331d} R. Foot, H. N. Long and Tuan A. Tran, Phys. Rev. D {\bf  50}, 34 (1994). 
\bibitem{331e} H. N. Long, Phys. Rev. D {\bf  54}, 4691 (1996). 
\bibitem{QQ1} F. Pisano and V. Pleitez, Phys. Rev. D {\bf  51}, 3865 (1995). 
\bibitem{QQ2} F. Pisano, Mod. Phys. Lett. A {\bf  11}, 2639 (1996). 
\bibitem{QQ3} A. Doff and F. Pisano, Mod. Phys. Lett. A {\bf  14}, 1133 (1999). 
\bibitem{QQ4} A. Doff and F. Pisano, Phys.Rev. D. {\bf  63}, 097903 (2001). 
\bibitem{QQ5} C. A. de S. Pires and O. P. Ravinez, Phys. Rev. D. {\bf  58}, 035008 (1998). 
\bibitem{QQ6} C. A. de S. Pires, Phys. Rev. D {\bf  60}, 075013 (1999). 
\bibitem{QQ7} P. V. Dong and H. N. Long, Int. J. Mod. Phys. A {\bf  21}, 6677 (2006).
\bibitem{QQ8} Adrian Palcu, Mod. Phys. Lett. A{\bf  24}, 2175 (2009).
\bibitem{Das} Prasanta Das and Pankaj Jain, Phys. Rev. D {\bf 62}, 075001 (2000).  
\bibitem{331x}  A. Doff, Phys. Rev. D {\bf 76}  037701 (2007). 
\bibitem{331y}  A. Doff, Phys. Rev. D {\bf 81}, 117702 (2010).
\bibitem{331z}  A. Doff and  A. A. Natale,  Int.J.Mod.Phys. A {\bf 27}  1250156 (2012).
\bibitem{331w}  A. Doff and  A. A. Natale,  Phys.Rev. D{\bf 87} 095004  (2013).  
\bibitem{tcb} S. Dimopoulos and L. Susskind, Nucl. Phys. B {\bf 155} , 237 (1979).
\bibitem{tce} E. Eichten and K. Lane, Phys Lett B {\bf 90}, 125 (1980).  
\bibitem{tc1} E. Farhi and L. Susskind,  Phys. Rept. {\bf 74},  277 (1981). 
\bibitem{tc2} C. T. Hill and E. H. Simmons, Phys. Rept. {\bf 381}, 235 (2003) [Erratum-ibid. {\bf 390}, 553 (2004)]. 
\bibitem{tc3} F. Sannino, {\it hep-ph/0911.0931 }, Lectures presented at the 49th Cracow School of Theoretical Physics. Conformal Dynamics for TeV Physics and Cosmology,  Cracow, Nov , 2009.  
\bibitem{tc4}K. Lane, {\it Technicolor 2000 }, Lectures at the LNF Spring School in Nuclear, Subnuclear and Astroparticle Physics, Frascati (Rome), Italy, May 15-20, 2000.
\bibitem{walk1} B. Holdom, Phys. Rev. D {\bf 24}, 1441 (1981) .
\bibitem{walk2} B. Holdom, Phys. Lett. B {\bf 150}, 301 (1985). 
\bibitem{walk3} T. Appelquist, D. Karabali e L. C. R. Wijewardhana, Phys. Rev. Lett. {\bf 57}, 957 (1986).
\bibitem{walk4} T. Appelquist and L. C. R. Wijewardhana, Phys. Rev. D {\bf 36}, 568 (1987). 
\bibitem{walk5} T. Appelquist, M. Piai, and R. Shrock, Phys. Rev. D{\bf 69}, 015002 (2004).
\bibitem{walk6} T. Appelquist, M. Piai and R. Shrock, Phys. Lett. B {\bf 593} , 175 (2004).
\bibitem{walk7} T. Appelquist and R. Shrock,  Phys. Rev. Lett. {\bf 90}, 201801-1 (2003).
\bibitem{walk8} T. Appelquist and R. Shrock, Phys. Lett. B {\bf 548} , 204 (2002).
\bibitem{walk9} M. Kurachi, R. Shrock  and K. Yamawaki,  Phys. Rev. D {\bf 76}, 035003 (2007).
\bibitem{peskin1} M. E. Peskin and T. Takeuchi, Phys. Rev. Lett. {\bf 65}, 964 (1990)
\bibitem{peskin2} M. E. Peskin and T. Takeuchi, Phys. Rev. D {\bf 46}, 381 (1992).
\bibitem{lane2} K. Lane,  Proceedings High energy physics, vol.2, 543-547, Glasgow 1994; hep-ph/9409304. 
\bibitem{sannino1}  F. Sannino and K. Tuominen, Phys.  Rev. D {\bf 71}, 051901 (2005). 
\bibitem{sannino2}  R. Foadi, M. T. Frandsen, T. A. Ryttov and F. Sannino, Phys. Rev. D {\bf 76}, 055005 (2007).
\bibitem{sannino3} T. A. Ryttov and F. Sannino, Phys. Rev. D {\bf 78}, 115010 (2008).
\bibitem{us1} A. Doff, A. A. Natale and P. S. Rodrigues da Silva, Phys. Rev. D {\bf 77}, 075012 (2008).
\bibitem{us2} A. Doff, A. A. Natale and P. S. Rodrigues da Silva, Phys. Rev. D {\bf 80}, 055005 (2009). 
\bibitem{us3} A. Doff, E. G. S. Luna and A. A. Natale, Phys. Rev. D {\bf 88}, 055008 (2013).
\bibitem{twoscale} A. Doff and A. A. Natale, Phys. Lett. B {\bf 748}, 55 (2015).
\bibitem{331-4f} A. Doff, Eur. Phys. J. C. {\bf 76}, 33 (2016).
\bibitem{us4} A. Doff and A. A. Natale, Phys. Lett. B {\bf 771}, 392 (2017). 
\bibitem{ardn}A. C. Aguilar , A. Doff and A. A. Natale,  hep-ph/1802.03206.
\bibitem{Craig} Craig D. Robertz and Bruce H. J. McKellar,  Phys. Rev. D  {\bf 41}, 672 (1990).
\bibitem{Pagels} H. Pagels and S. Stokar,  Phys. Rev. D {\bf 20}, 2947 (1979).
\bibitem{lane1} K. Lane,   Phys. Rev. D  {\bf 10}, 2605 (1974).
\bibitem{ope} H. D. Politzer, Nucl. Phys. B {\bf 117}, 397 (1976).
\bibitem{She5} She-Sheng Xue, Mod.Phys.Lett.A {\bf 14}, 2701-2708,(1999).
\bibitem{She4} She-Sheng Xue, Phys. Lett. B {\bf 398}, 177-186,(1997).
\bibitem{She3} She-Sheng Xue, Phys. Lett. B {\bf 721}, 347-352 (2013). 
\bibitem{She2} She-Sheng Xue,  J. High Energ. Phys. {\bf 72} (2016). 
\bibitem{She1} She-Sheng Xue,  Phys. Rev. D {\bf 93}, 073001 (2016). 
\bibitem{alex1} Alex G. Dias and V. Pleitez, Phys.Rev. D {\bf 80}, 056007 (2009). 
\bibitem{alex2} Alex G. Dias, J.C. Montero and  V. Pleitez,  Phys.Rev. D{\bf 73} 113004 (2006). 
\bibitem{alex3} A. G. Dias, R. Martinez and V. Pleitez, Eur. Phys. J. C{\bf 39} 101-107 (2005).
\bibitem{alex4} Alex Gomes Dias, Phys.Rev. D{\bf 71}, 015009 (2005).  

\end {thebibliography}
\end{document}